\documentclass[conference]{IEEEtran}
\IEEEoverridecommandlockouts
\usepackage{cite}
\usepackage{amsmath,amssymb,amsfonts}
\usepackage{algorithmic}
\usepackage{graphicx}
\usepackage{textcomp}
\usepackage{xcolor}
\usepackage{tikz}
\usepackage{pgfplots}
\usepackage{todonotes}
\usepackage{soul}

\pgfplotsset{compat=1.18}
\def\BibTeX{{\rm B\kern-.05em{\sc i\kern-.025em b}\kern-.08em
    T\kern-.1667em\lower.7ex\hbox{E}\kern-.125emX}}
\begin{document}

\title{How Critical is Site-Specific RAN Optimization?  5G Open-RAN Uplink Air Interface Performance Test and Optimization from Macro-Cell CIR Data\\
\thanks{This work was supported in part by the National Telecommunications and Information Administration (NTIA) under the Federal Award ID 51-60-IF004 (PWSCIF NOFO1).}
}

\author{Johnathan Corgan${}^{1,*}$, Nitin Nair${}^{1,*}$, Rajib Bhattacharjea${}^1$, Wan Liu${}^1$, Serhat Tadik${}^{1,2}$, Tom Tsou${}^1$, Timothy J O'Shea${}^1$ \\
\footnotesize \begin{tabular}{cccccc}
${}^1$Deepsig, Inc. & ${}^2$ Georgia Institute of Technology & ${}^*$Authors contributed equally\\
\end{tabular}
}

\maketitle

\begin{abstract}
In this paper, we consider the importance of channel measurement data from specific sites and its impact on air interface optimization and test.  Currently, a range of statistical channel models including 3GPP 38.901 tapped delay line (TDL), clustered delay line (CDL), urban microcells (UMi) and urban macrocells (UMa) type channels are widely used for air interface performance testing and simulation.  However, there remains a gap in the realism of these models for air interface testing and optimization when compared with real world measurement based channels. To address this gap, we compare the performance impacts of training neural receivers with 1) statistical 3GPP TDL models, and 2) measured macro-cell channel impulse response (CIR) data. We leverage our OmniPHY-5G neural receiver for NR PUSCH uplink simulation, with a training procedure that uses statistical TDL channel models for pre-training, and fine-tuning based on measured site specific MIMO CIR data. The proposed fine-tuning method achieves a 10\% block error rate (BLER) at a 1.85 dB lower signal-to-noise ratio (SNR) compared to pre-training only on simulated TDL channels, illustrating a rough magnitude of the gap that can be closed by site-specific training, and gives the first answer to the question ``how much can fine-tuning the RAN for site-specific channels help?''
\end{abstract}

\begin{IEEEkeywords}
Neural Receiver, Site-Specific Optimization, 5G NR, 5G Advanced, 6G, OpenRAN, Air Interface, Uplink, RAN Digital Twin, Test and Measurement, Channel Simulation
\end{IEEEkeywords}

\section{Introduction}
Conventional model-driven algorithms used in wireless communications systems are designed to work on a set of closed-form statistical models that approximate the behavior and effects within a complex system.  For example, channel models such additive-white-Gaussian noise (AWGN) and the Rayleigh fading model capture much of the rich dynamics of the wireless channel, but do not capture the effects of all relevant phenomena (e.g., spatial and angular properties, co-channel interference from other users, or amplifier or other hardware non-linearities), and are not specific to any particular location or relative geometry among the transmitter, receiver, and scatterers in the environment. More sophisticated models, such as the 3GPP TR 38.901 \cite{3gppTR38901}, do consider the relative geometries of the transmitter and receiver, antenna patterns, and if line-of-site (LOS) or non-line-of-site (NLOS) conditions exist between the transmitter and receiver.  Conventional wireless communications system algorithm design is then typically based on the same simplifying assumptions that underlie the channel models, and algorithms are typically evaluated against those same channel models themselves. Algorithms designed in this way have come a long way and are generally robust enough to operate well in the real world despite being designed and tested against the approximations. However, there are universally cases of `model deficit' where the channel model and simulation model do not capture all effects encountered in the real world, and which change the underlying channel conditions and assumptions of the system modeling and design processes, leading to degraded wireless communications system performance.

Recently, through the growth of AI/ML and deep learning methods, platforms, and tools, the use of vastly more data in addition to models has become feasible in both signal processing solutions and modeling of environments.   To this end, there has recently been significant work on using RAN Digital Twin (RAN-DT), and data-driven propagation models \cite{hoydis2023sionnartdifferentiableray} to enable site-specific learning.   

\begin{figure*}[ht]
    \centering
    \includegraphics[width=.9\linewidth]{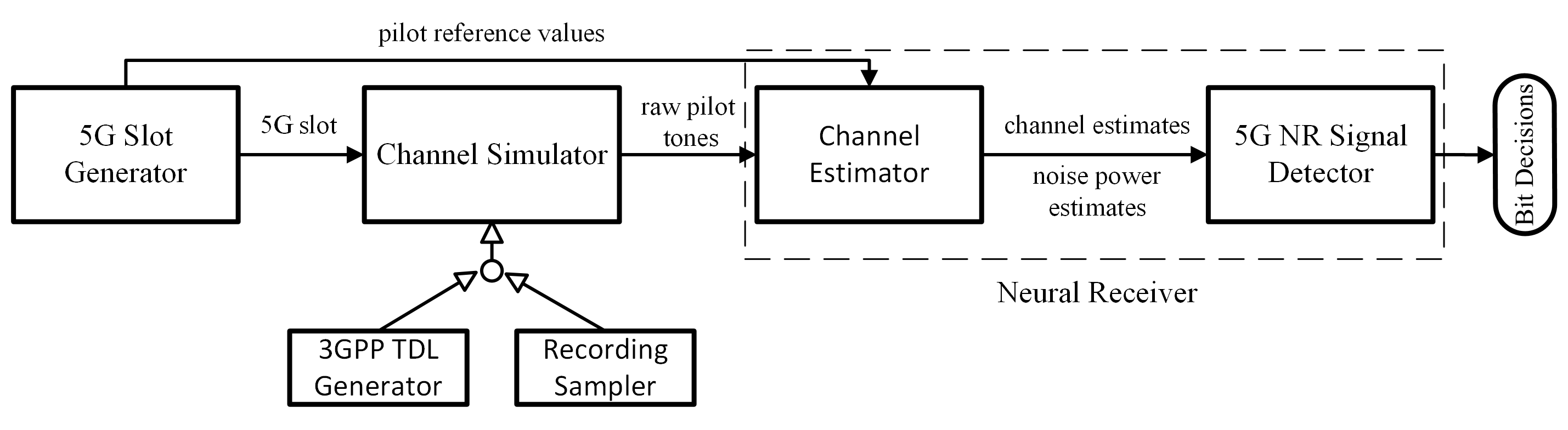}
      \caption{5G neural receiver training and evaluation block diagram}
    \label{fig: block_diagram}
\end{figure*}

This paper contains the results of combining a novel site-specific macro-cell channel measurement methodology with the training and evaluation of a neural receiver under the measured channel conditions. As a baseline, our neural receiver is first pre-trained on simulated 5G uplink slots passed through a 3GPP TDL channel simulator (see Figure \ref{fig: block_diagram}), and is then fine-tuned on site-specific over-the-air channel measurements collected passively from commercial downlink reference signals.

Our main contributions in this work are threefold:
\begin{itemize}
    \item We describe a new data collection methodology for over-the-air macro-cell channel measurements collected opportunistically from downlink reference signals.
    \item We illustrate that captured CIR data can be used for OpenRAN Air Interface test and measurement to obtain realistic performance measurement for L1 functions.
    \item We show that pre-training a neural receiver using a 3GPP TDL channel model, followed by fine-tuning with site-specific CIR data achieves a 10\% Block Error Rate (BLER) at a 1.85 dB lower signal-to-noise ratio (SNR) compared to pre-training alone.
\end{itemize}

The remainder of the paper is organized as follows. Section \ref{sec: related_work} describes the relevant works studied in the literature. Section \ref{sec: methodology} describes the measurement system \& campaign along with the 5G NR neural receiver pre-training, fine-tuning, and evaluation. Section \ref{sec:results} shares the results, and Section \ref{sec: conclusion} discusses them along with the limitations and concludes the paper.

\section{Related Work} \label{sec: related_work}
Designing in-line physical layer algorithms for use in wireless communications systems via deep learning is a relatively new area of research, only emerging in the literature in approximately the last 8 years. The key insight is to re-imagine an entire wireless communications link as an autoencoder, or a series of neural networks which reconstructs its input at the output, in which the information that is transmitted into the system needs to be received with minimal bit or block errors. In this context, the transmitter and receiver algorithms such as modulation, coding, channel estimation, demodulation, and decoding can be replaced with neural networks that are trained in such a way as to form an autoencoder in the presence of the unknown channel \cite{ae_comms_siso_oshea}. These ideas have been extended to a multi-input multi-output (MIMO) configurations by incorporating multiple antennas at both the transmitter and receiver \cite{ae_comms_mimo_oshea}. While these approaches showed promise, the end-to-end autoencoder approach requires a differentiable channel model for backpropagation, which can be difficult after deployment. This limitation has led to the adoption of channel-agnostic end-to-end systems that leverage stochastic perturbation techniques \cite{spsa_chan}, generative channel models, or other statistical or geometric channel models, replacing the traditional fixed formulations of channel models \cite{cgan_chan, vgan_chan_oshea, wgan_chan, hoydis2023sionnartdifferentiableray}. Generative channel model approaches have also been extended to MIMO and massive MIMO channel configurations \cite{cgan_chan_mimo, gan_chan_mmimo}.

The autoencoder based approach is interesting in the context of a new ``AI-Native'' waveform design containing new modulation, coding, pilot/frame structure, etc (e.g. 6G); however, in the current generation of wireless communications systems such as WiFi, 4G and 5G, the specifications contain a rigidly defined frame structure, modulation, reference signals and coding scheme - lending themselves to a classical fixed and model-driven transmitter. In this setting, learnable receivers have been proposed \cite{chan_est_nr1} for use in systems that retain traditional transmit algorithms that define the physical layer while incorporating learnable parameters to enhance receiver adaptability and performance. We call such a receiver a {\em neural receiver} because almost all implementations in the literature of fully-learnable receiver algorithms are based on neural networks of some form.

Neural receivers are typically designed using data-driven or model-driven methodologies \cite{nr2}. In model-driven architectures, there are two common configurations: some designs use separate neural network modules for each part of the traditional receiver chain such as channel estimation \cite{chan_est_nr2}, demapping \cite{demapping_nn}, and decoding \cite{decoding_nn} while others separate the receiver architecture into neural networks with combined functionalities such as signal detection and channel estimation \cite{nr1, chan_est_signal_det_nr, nr2}.

The training of neural receivers is usually performed with either simulated synthetic datasets with certain channel models or over-the-air datasets with real channel distortion effects. Typical channel models used for data simulation include IEEE 802.11a TGn multipath fading channel \cite{nr1}, a geometric scattering channel \cite{chan_est_nr2}, 3GPP tapped delay line (TDL) channel \cite{chan_est_signal_det_nr}, the wireless world initiative new radio II (WINNNERII) and Stanford University Interim (SUI) models \cite{nr2}. 

Channel sounding measurements are typically made using transmitter and receiver devices with a channel sounding waveform \cite{wili_dataset} or by capturing downlink reference signals that can be post-processed into channel responses. Two examples from the literature put the receiver on a UAV \cite{lte_meas} or on a pedestrian walking in an outdoor area measuring commercial LTE tower downlink signals with an antenna \cite{pretrain_chan}. These type of data collection techniques capture large volumes of IQ data, and post-process them into a much smaller volume of channel responses. The limitation of these approaches is that most of the captured IQ is not needed for channel estimation; indeed, the majority of the captured data would be noise and downlink data addressed to handsets, not the pilot tones useful for channel estimation. In such a system, most of the finite storage volume of the receiver is filled with data that is ultimately of no use for channel measurement, and so the amount of time these systems can operate to collect data is much less than a system that processes the downlink signal in real-time and only stores the pilot information. We draw attention to this distinction because we have used the real-time demodulation technique in this work, which allows us to continuously collect channel measurements for hours at a time without being constrained by the limitations of the data storage medium.

\section{Methodology} \label{sec: methodology}

\subsection{Measurement System and Measurement Campaign}

Channel measurements were taken passively using ambient commercial LTE downlink signals that are nearly ubiquitous in urban areas. The LTE specification requires cell-specific reference signals in the form of downlink modulation reference symbol (DMRS) pilots to be in fixed locations in the downlink OFDM grid, and fully populated for each slot. These are used by a UE (e.g., a handset) to perform channel estimation and channel equalization when decoding PBCH, PDCCH, PDSCH, etc.  These DMRS can be used as a sparse channel sounding waveform in that they are known sequences, and can be used by a receiver to continuously estimate the channel between the eNodeB transmit antennas and the receiver. These channel estimates are specific to the band of operation of the LTE downlink signal, {\em not} the wireless communication method itself. The channel estimates may therefore be used more broadly to simulate propagation of any type of signal operating in that band with a similar geometry. In this work, we leverage these measured channel responses to validate the real-world performance of a 5G NR OpenRAN uplink air interface, and to train a corresponding 5G NR neural receiver to improve performance for the site.

LTE downlink DMRS data was collected at several frequencies; Table \ref{tab:measurement_parameters} provides the LTE parameters for the measurements we focus on for this paper. The receive antenna is a commercially available, adjustable telescoping monopole antenna magnetically mounted on a car roof. This antenna is connected to a 710-850 MHz bandpass filter \cite{minicircuits_VBFZ780S}, which is connected to a low-noise amplifier (LNA) with 0.38 dB typical noise figure \cite{minicircuits_ZX60P33ULN}. The LNA output is connected to a Ettus USRP B210 software-defined radio operating at a sample rate of 30.72 MS/s. The B210 also uses a GPS disciplined oscillator (GPSDO) module, so a separate GPS antenna is also mounted on the vehicle roof and a cable connects it to the GPS port of the B210.

\begin{table}[htbp]
\centering
\caption{Measurement parameters.}
\begin{tabular}{lcl}
\hline
\textbf{Parameters} & \textbf{Values} & \textbf{Descriptions} \\
\hline
$f_c$ & 751 & Carrier frequency [MHz] \\
$BW$ & 9 & Measurement bandwidth [MHz] \\
$N\_RB$ & 50 & Number of resource blocks \\
$N\_Tx$ & 4 & Number of transmitters \\
$N\_Rx$ & 1 & Number of receivers \\
\hline
\end{tabular}
\label{tab:measurement_parameters}
\end{table}

Data collection proceeds by driving the vehicle with the aforementioned hardware being controlled by a custom software application we call {\em hcapture} that runs on a laptop inside the vehicle. It communicates with the B210 via a USB 3.0 interface using the standard UHD driver. The {\em hcapture} software synchronizes to LTE downlink signals using correlation with synchronization sequences (PSS and SSS). Successful PSS and SSS detection provides the physical cell identity (PCI) and allows for the generation of a correct DMRS sequence and demodulation of the PBCH/MIB with high level cell information. Synchronization also allows {\em hcapture} to obtain precise LTE frame timing, and extraction of DMRS tones in the downlink OFDM grid for a given PCI, so channel estimates and raw received pilots can be stored. This is a relatively minimal amount of recorded data needed to estimate the channel response from the tower to the receiver and maintain high level sector broadcast and PCI information.  {\em hcapture} uses a minimum-mean-squared error channel estimation \cite{proakis_salehi_digital_communications_2020} process to recover these channel estimates across all active sub-carriers. GPS parameters such as position, fix accuracy, and others, are also queried from the aforementioned GPSDO module and logged with every record (e.g. per-slot).

\begin{figure}[ht]
    \centering
    \includegraphics[width=.9\linewidth]{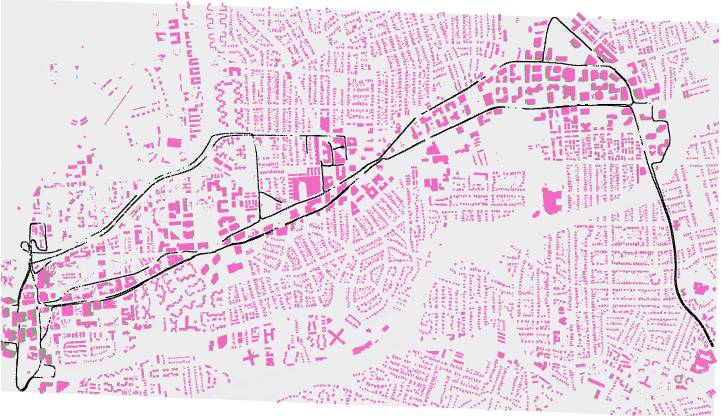}
      \caption{An example drive-map of {\em hcapture} channel sounding in Arlington, VA area. The map is an overhead view, with pink structures representing building geometry and the black points and lines representing the drive route along roads.}
    \label{fig:map}
\end{figure}

Data collection campaigns were conducted in Atlanta, GA, Arlington, VA, and San Jose, CA, in the first half of 2024, with the same measurement parameters and at driving speeds typical to urban and highway driving scenarios (from about 0 to 70 mph). A sample drive route is depicted in Figure \ref{fig:map}. The results of this paper focus on the Atlanta dataset, which was collected on February 24, 2024 over a drive of about 42 minutes. Data was captured from 23 unique PCIs, and a total of 87,989 records were captured, with each record containing 12 measurements of the channel, possibly slightly offset in time from each other, and originating from 4 independent antennas on the tower.

\subsection{Neural Receiver Pre-Training, Fine-tuning, and Evaluation}
Figure \ref{fig: block_diagram} shows the system block diagram that governs how the 5G NR neural receiver is pre-trained on synthetic channel realizations and fine-tuned on measured data from the measurement campaign of the previous section. Training slots are generated according to the 5G NR specification with both data and pilot symbols present, with slots of a fixed modulation and coding scheme (MCS) and random transport-block (TB) bit contents. The channel simulator applies the MIMO channel impulse response from {\em hcapture} to the slot, using a channel realization that was seen over-the-air during measurement, and Gaussian noise is added at a specific relative noise power to achieve a desired signal-to-noise ratio (SNR). The selection of which channels are used for training and testing can vary depending on how ``site-specific'' the desired evaluation process is. For instance, it may be desirable to test a single PCI's locality performance using only samples from the PCI, or it may be of interest to sample only channel responses from a region of portion of a city more broadly. Finally, the slots are received by the neural receiver which performs the functions of channel estimation, equalization, demodulation and demapping required to generate soft-bit output corresponding to log-likelihoods of the original data which can be decoded. The bit error rate (BER) and block error rate (BLER) is measured at different SNR operating points for the receiver while using random data and samples of real channel responses. We call this a {\em BLER sweep}. A ``passing'' BLER such as 10\% is chosen as a target, corresponding to a BLER target level which might be used by a 5G MAC scheduler, and an intersection is interpolated between measured BLER values to determine a ``passing'' SNR for the BLER sweep. This can be repeated for any number of receiver algorithms to obtain corresponding sensitivity performance for passing SNR, and relative SNRs can be compared in order to inspect relative sensitivity of different receiver algorithms. Neural receivers that can achieve this 10\% error rate at lower SNRs are considered better because they can successfully operate in noisier channels, or equivalently, at longer ranges between transmitter and receiver. For training the neural receiver, the difference between the decoded bits and the known transmitted bits is used to define a loss function; then a standard variant of gradient descent optimization is used to update the weights of the neural receiver to reduce the value of the loss function in a process called training. We consider our baseline neural receiver as one that is trained against channel realizations drawn from TDL scenarios, and compare that against a fine-tuned neural receiver that is pre-trained on TDL scenarios but also undergoes fine-tuning (training with a small learning rate) on the Atlanta channel captures as described in the previous section. The TDL scenarios are specified in \cite{3gppTR38901} and our pre-training uses a mix of TDL-B and TDL-C channels with delay spreads ranging from 10 to 600 nanoseconds and Doppler ranging from 20-400 Hz, a pre-training configuration which has been found to generalize reasonably well on other test sets. We also present results for a conventional, non-ML algorithm (MMSE receiver) so that the reader can see how neural receivers generally compare favorably to conventional algorithms. Both the pre-trained baseline and fine-tuned neural receivers are then evaluated by performing the BLER sweep process, with the evaluation channel responses themselves drawn from the measured dataset. We evaluate the models with different modulation and coding schemes (MCS) on the user data payloads according to Tables 5.1.3.1-2 of 3GPP TS 38.214 \cite{3gppTR138214}.

Finally, we note that uplink performance is critical in the deployment of any 5G system, including OpenRAN systems.  Because uplink transmit powers are much more limited than downlink transmit powers, coverage area is generally limited by uplink sensitivity (i.e., the lowest RSRP possible for a specific MCS under varying conditions) and total uplink throughput or capacity per-user or per-sector is also limited by how uplink sensitivity in terms of how much spectral efficiency (i.e., bits/s/Hz) can be achieved using higher MCS allocations with low-BLER consistently for cell users.  This justifies why we chose to evaluate the SNR at which BLER performance achieves an acceptable level for receiver algorithms, simulating the impact on ultimate RSRP sensitivity performance. Being able to realistically simulate and measure this under real world conditions is important for optimizing, comparing, or choosing the best receiver approach to maximize value and performance when deploying a 5G system.

\section{Results}\label{sec:results}
\begin{figure*}[ht]
  \centering
  \includegraphics[width=.5\linewidth]{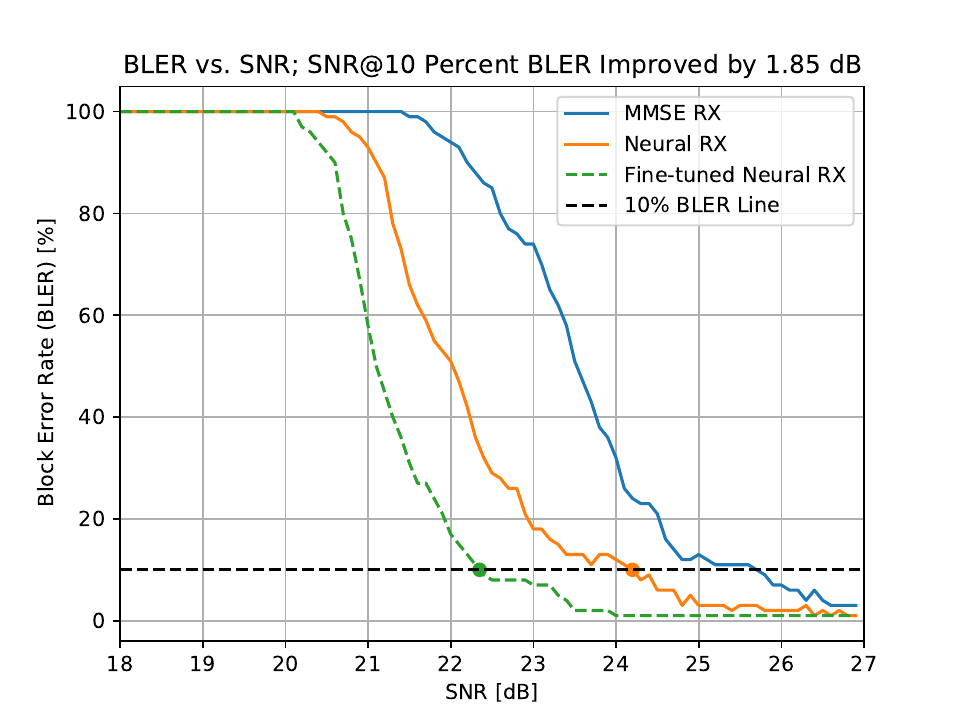}
    \caption{Comparison of the baseline neural receiver vs. one fine-tuned on site-specific data (MCS 27: 256QAM with a rate 0.93 code)}
  \label{fig:result}
\end{figure*}

Figure \ref{fig:result} illustrates key BLER sweep results showing the improvement from fine-tuning on local channel measurements, where both baseline and fine-tuned models are evaluated on the same measured channel distribution. The largest impact was seen on the illustrated high MCS (27) condition, corresponding to data symbols with 256-QAM and a rate-0.93 code \cite{3gppTR138214}. The key result here is that sensitivity (SNR) improved by 1.85 dB for a typical target BLER (10\%) for the fine-tuned model as compared to the baseline model, indicating that this MCS could be used in a wider range of cell conditions increasing capacity throughout the coverage area. The full shift in the BLER curve can also be seen, showing improvement at lower and higher BLER operating points as well.  Evaluation under MCS 20 (64-QAM with a rate-0.55 code) was also considered but exhibited less benefit in this case, showing that under this experiment, the biggest fine-tuning benefit was seen at high MCS. Note that the BLER sweep results for a conventional algorithm, MMSE, is also presented as a reference that demonstrates that both neural receivers outperform conventional receivers, and serves as a check that the neural receiver training processes converged.

\section{Discussion, Limitations, and Conclusion} \label{sec: conclusion}
We have presented a first-of-its-kind analysis of site-specific fine-tuning of neural receivers on real world measurement based macro-cell channel data. Site-specific data have been collected using a novel channel measurement system. The site-specific data were used to fine-tune a pre-trained neural receiver for 5G NR PUSCH. In the best case, corresponding to high MCS values that use larger constellations and less error correction, the fine-tuned model showed it could operate in nearly 2 dB lower SNR relative to the TDL pre-trained model while maintaining a 10\% block error rate. At lower MCS values (lower order QAM and spectral efficiency), the site-specific fine-tuning provided less improvement, and further investigation is required in order to explore multi-MCS fine-tuning strategies. Significant open issues remain on capturing training into generative models, deciding "how site specific" a model and its training should be, and generally optimizing the methodology to obtain the best possible fine-tuning performance; all of these items will be addressed in future work.

\bibliographystyle{IEEEtran}
\bibliography{biblio}

\end{document}